\documentclass{article}
\usepackage{graphicx}
\usepackage{latexsym}
\usepackage{amsmath}
\usepackage{amssymb}
\usepackage{tensor}
\newcommand{\N}{\tensor{N}}

\newcommand{\be}{\begin{equation}}
\newcommand{\ee}{\end{equation}}
\newcommand{\bea}{\begin{eqnarray}}
\newcommand{\eea}{\end{eqnarray}}
\title{ Stochastic metric perturbations (radial) in  gravitationally collapsing 
spherically symmetric relativistic star  }
\author{Seema Satin  \\
\small Dept. of Physical Sciences , Indian Institute for Science Education and
Research , Mohali, India.\\
\small satin@iisermohali.ac.in}
\vspace{10pt}
\date{ }
\begin{document}
\maketitle
\begin{abstract}
 Stochastic perturbations (radial) of a spherically symmetric  relativistic
 star, modeled by a perfect fluid in comoving coordinates for the collapse
scenario are worked out using the
 classical Einstein- Langevin equation, which has been proposed recently. 
The solutions are in terms of  perturbed metric potentials and their
 two point correlation. For the case worked out here, it is interesting 
to note that the two perturbed metric potentials have same magnitude, while the
potentials themselves are in general independent of each other. Such a 
treatment is useful for building up basic theory of non-equilibrium and
 near equilibrium statistical physics for collapsing stars, which should be
 of interest towards the end states of collapse. Here we discuss the first
simple model, that of non-rotating spherically symmetric  dynamically
 collapsing relativistic star. This paves way to further research on
 rotating collapse models of isolated as well as binary configurations on 
similar lines . Both the radial and non-radial perturbations with stochastic
effects would be  of interest to asteroseismology, which encompassed the future
plan of study.
\end{abstract}
\section{Introduction}
Perturbations of relativistic stars are of interest to studies in 
asteroseismology \cite{bardeen,fried} and has recently gained more importance
 in the context of 
gravitational waves, emitted by collapsing stars or binary mergers \cite{abbot}. 
Nevertheless, perturbations of such configurations have a wider range of
 relevance \cite{chandra1,chandra2,chandra3}. 
 The oscillations of stars, both radial
and non-radial have been studied and mathematical formulations of the
 theory has a long standing history \cite{chandra4,chandra5,kipnotes}. These
 oscillations enables one to
 study the interior of the stars, which for the more relativistic cases, 
is an active area of research. The equations of state of the matter content of
 a  compact star, towards the end state of collapse is open to investigations.
 Also the  dynamics of the interior of such a collapsing star is of interest, 
and ways to determine this still awaits theoretical as well as observational
 developments.
The metric perturbations to the spacetime geometry are often treated 
deterministically, and modes of oscillations addressed for a given 
compact system under investigation. This is done via the perturbed Einstein's
equations and their solution. 
 
In this article we attempt to show  how stochastic effects in the matter
content of relativistic star give rise to metric perturbations with induced
 stochasticity. The aim of such an approach is to set up a theoretical
 framework for doing non-equlibrium statistical analysis of a relativistic
star, which should be of interest and more relevant in observational aspects,
 towards the end states of dynamical 
collapse.  Here we treat a very simple non-rotating geometry, to show
that an analytical method can be developed for the same, though this article
is restricted to  radial perturbations only . Such a case for static stellar
 configuration has been recently worked out \cite{seema}, where we have
 proposed a classical Einstein
 Langevin equation in the context. This article takes forward the same theme,
 for
 a collapsing star in comoving coordinates and is the next step in the 
development. On the lines of this dynamical collapse, we will in future
 attempt to consider rotating geometry and non-radial perturbations which will
 takes us ahead on the program for more realistic cases. 

An interesting feature in using this formalism is that, the relativistic
 Cowling approximation is ignored \cite{fried,lee}, hence it is exact solution
 to Einstein's equation, along with added stochastic effects which are seen to 
arise  as a result of statistical
properties of matter content in the interior of the star. As is apparent from
 the results obtained form our preliminary calculations, it is not the
fluctuations of the pressure and density at a given spacetime point that show up
 playing a direct role in the metric perturbations, but the covariances of
 pressure and density inside the star at separated points, which induce these.
 Hence, one can see that the cowling approximation  does not find a way 
directly in this framework, and
we get results from taking the perturbed equations in their complete form. 
In this  analysis  the averages of the metric perturbations $<h^{ab}(x)>$, are
 vanishing, which does in fact fall in line with  cowling approximation. Thus
one can view it as cowling approximation been hidden in the stochasticity,
 while we see the real effects  of randomness in terms 
of two point correlations of the metric, and covariances of fluid elements. 
This will become clearer, when we deal with non-radial perturbations in which
 case the Cowling approximation is usually applied.

In the specific example that we take up in this article the noise in the
 system or the source of stochasticity are the pressure and density 
covariances in the perfect fluid composing a compact object. Thus the
 randomness in the stress tensor induces the statistical effects in the
 perturbed geometry as we see below.

This is the most simplified model, more involved cases
being those of dissipative and anisotropic fluids \cite{herra1,herra2,ban} in
 a collapsing scenario
 which form  realistic matter fields towards end states of collapse in
 highly compact configurations.This can be dealt  
in future work, the common feature  carried forward being the
fluctuations of the stress tensor for anisotropic and disspative fluids,
would be defined in terms of, say, flucutations of heat flux in the interior
of the star as it collapses. However, it
may be difficult to get analytical solutions for those cases and, one may
have to succumb to  numerical solutions. The below formalism sets stage for
such cases, where the physical results would encompass  dependence 
of the metric perturbations on statistical properties of dissipative fluids  
which would show up explicitly and probably a fluctuation- dissipation 
 relation in the system addressed.   

\section{ The collapse model  with stochastic effects }
 
For a non-rotating spherically symmetric relativistic star in comoving 
coordinates given by 
\be
ds^2 = - e^{2 \nu(t,r)} dt^2 + e^{2 \psi(t,r)} dr^2 + R^2(t,r) d \Omega^2
\ee 
 with a perfect fluid stress tensor
\be \label{eq:stress}
T^t_t = - \rho(t,r), T^r_r =  T^\theta_\theta = T^\phi_\phi = p(t,r)
\ee
 the Einstien's tensors read
\begin{eqnarray}
G^t_t & = & - \frac{F'}{R^2 R'} + \frac{ 2 \dot{R} E^{- 2 \nu}}{R R'} (\dot{R}'
- \dot{R} \nu' - \dot{\psi} R') \\
G^r_r & = & - \frac{\dot{F}}{R^2 \dot{R}} - \frac{2 R'}{R \dot{R}} e^{-2 \psi}( 
\dot{R}' - \dot{R} \nu' - \dot{\psi} R') \\
G^t_r & = & - e^{2(\psi-\nu)} G^r_t = \frac{ 2 e^{-2 \nu}}{R} ( \dot{R}' -
\dot{R} \nu' - \dot{\psi} R') \\
G^\theta_\theta & = & G^\phi_\phi = e^{-2 \psi}[ ( \nu'' + \nu'^2 - \nu'
 \psi')R + R'' + R' \nu' - R' \psi'] \nonumber \\
& &  - \frac{e^{-2 \nu}}{R} [ ( \ddot{\psi} - 
\dot{\psi}^2 - \dot{\nu} \dot{\psi})R + \ddot{R} + \dot{R} \dot{\psi} - 
\dot{R} \dot{\nu} ]
\end{eqnarray}
where $F$ is defined by the equation
\be
e^{2 \psi} = (1 + e^{-2 \nu} \dot{R}^2 - \frac{F}{R})^{-1} R'^2
\ee
The perfect fluid stress tensor can be treated as random variable, while
the randomness can be introduced in the system by internal or external
 influences. The fluid particles thus undergo random collisons, which may 
 not be of thermal origin, since for 
compact objects like neutron stars we deal with non-thermal effects, that of
 quantum origin where the pressure in the star is due to neutron degeneracy 
and the like. The fluctuations of the macroscopic fluid variables may thus
partially capture the quantum effects in the dynamical system.  The external
 influences that can give rise to randomness, may be
 mechanical
in origin like the accretion of matter or implosion effects of the core of the
star. Though here we  deal with a non-rotating system, these effects 
 are very much known to induce perturbations in the system. Usually these
 perturbation are treated deterministically. 
We introduce the randomness in a fashion, such that the 
system can be modeled by an Einstein Langevin equation \cite{seema}  
\be
G^{ab} [g+h](x) = T^{ab}[g+h](x) + \xi^{ab}[g](x)
\ee
where $\xi^{ab}(x)  = (T^{ab}(x) - <T^{ab}(x)>) $ is the Langevin noise
 defined by $<\xi^{ab}(x)> = 0$, satisfying $\nabla_a \xi^{ab}(x) = 0 $.
The two point correlation $<\xi^{ab}(x) \xi^{cd}(y) >  = N^{abcd}(x,y) $,
as discussed later , defines the noise. It is important to note that the
noise term $\xi^{ab}$ is defined on the background spacetime $g^{ab}(x)$ and
 not the
perturbed one. In fact it is this noise that is supposed to induce the
perturbations in the metric over the background.

The above equation can be written as
\be \label{eq:el1}
\delta G^{ab}(x) = \delta T^{ab}(x) + \xi^{ab}(x)
\ee
as the Einstein's equation balances the unperturbed part 
$G^{ab}[g] = T^{ab}[g]$ over the background geometry.
 
 The perturbed Einstein tensors take the following form, where we assume the 
model with the perturbation in the area radius  $ \delta R(t,r) = 0 $, for
 mathematical simplicity.
\begin{eqnarray}
\delta G^t_t  & =  & 2 \delta \nu e^{-2 \nu} \frac{\dot{R}}{R}
 ( \frac{\dot{R}}{R} + 2 \dot{\psi}) - 2 \delta \psi e^{-2 \psi}
 ( \frac{R'^2}{R^2}  \nonumber \\
& & - 2 \psi'\frac{R'}{R} + 2 \frac{R''}{R} ) - 2 \delta \dot{\psi}
 \frac{\dot{R}}{R} e^{-2 \nu} - 2 \delta \psi' \frac{R'}{R} e^{-2 \psi} 
\label{eq:gtt}\\
\delta G^r_r & = & 2 \delta \nu e^{-2 \nu} ( \frac{\dot{R}^2}{R^2} + 2 
\frac{\ddot{R}}{R} + 2 \dot{\nu} \frac{\dot{R}}{R}) - 2 \delta \psi e^{-2 \psi}
( \frac{R'^2}{R^2} + 2 \frac{R'}{R} \nu') \nonumber \\
& &  - 2 \delta \dot{\nu} \frac{\dot{R}}{R} e^{-2 \nu} + 2  \delta \dot{\psi}
 \frac{R'^2}{R \dot{R}} e^{-2 \psi} \label{eq:grr}  \\
\delta G^t_r & = & - \dot{R} \delta \nu' - \delta \dot{\psi} R' \label{eq:gtr}
\end{eqnarray}
and similar expression for $ \delta G^\theta_\theta $ and $\delta G^\phi_\phi$,
which we can use in the Einstein Langevin equation (\ref{eq:el1}) for the
geometry sector.
The perturbed stress tensor has non-zero components, $\delta T^t_t
= - \delta \rho, \delta T^r_r = \delta T^\theta_\theta = \delta T^\phi_\phi
= \delta p $ . For the stochastic term $\xi^{ab}$, we defined noise in the
following section.
\subsection{ Noise in the system} \label{sec:noise}
The noise term characterized by $\xi^{ab} $ in the Einstein Langevin equation
 is responsible for the induced stochasticity in the perturbed metic
potentials, as we would see further in this article. This is the
 Langevin noise which for the perfect fluid stress tensor as defined above
is given by $\xi^{ab} = 0 $ , while the two point correlation gives
 a non-zero contribution and defines randomness in the system as shown below.
The most general definition as proposed in \cite{seema} is
\begin{eqnarray}
N^{abcd}(x,x') & =&  <(T^{ab}(x) - <T^{ab}(x)>)(T^{cd}(x') - <T^{cd}(x')>)> 
\nonumber \\
& = & <T^{ab}(x) T^{cd}(x') > - <T^{ab}(x)><T^{cd}(x')> \nonumber \\
& = & Cov[T^{ab}(x), T^{cd}(x')]
\end{eqnarray}
which  in a slightly different way with the indices can be written as
\be
\N{^a_b^c_d}(x,x') = <\xi^a_b(x) \xi^c_d(x')> = Cov[T^a_b(x),T^c_d(x')]  
\ee
For the stress tensor given by (\ref{eq:stress}), the nonzero noise components
can be directly read off from the above definition as
\begin{eqnarray}
& & \N{^t_t^t_t}(t,r;t',r') =  Cov[\rho(t,r),\rho(t',r')], 
\N{^t_t^r_r}(t,r;t'r') = \N{^t_t^\theta_\theta}(t,r't'r') = \nonumber\\
& &  \N{^t_t^\phi_\phi}(t,r;t'r') = -Cov[\rho(t,r), p(t,r)] 
\nonumber\\
& & \N{^r_r^t_t}(t,r;t',r') = \N{^\theta_\theta^t_t}(t,r;t'r') =
 \N{^\phi_\phi^t_t}(t,r;t',r') = - Cov[p(t,r),\rho(t,r)] \nonumber \\
& & \N{^r_r^\theta_\theta}(t,r;t',r') = \N{^r_r^\phi_\phi}(t,r;t',r') = 
\N{^\theta_\theta^\phi_\phi}(t,r;t',r') =
 \N{^\phi_\phi^\theta_\theta}(t,r;t'r') \nonumber \\
& &  = Cov[p(t,r),p(t',r')]
\end{eqnarray}
\subsection{Solution of Einstein Langenvin equation}
It is interesting to note here that for our model, in which we assume
 $\delta R(t,r) = 0 $ , equation (\ref{eq:gtr}) gives , (as $\delta T^t_r = 0$
simlarly $\xi^t_r = 0 $) 
\be  \label{eq:soltr}
\frac{\delta \dot{\psi}}{\dot{R}} = - \frac{\delta \nu' }{R'}
\ee 
Hence it follows that $\delta \psi =  - \delta \nu $. 
Thus the two perturbed potential are same , except for the difference in sign,
 even though there is no such restriction on the potentials themselves. This
is an interesting result in itself, which enables one to simlify
 the complexity of Einsteins' perturbed equations as well as  the Einstein
Langevin equation drastically, as we see further in this article. 

In view of the above relation between the  $\delta \psi $ and $\delta \nu$,
equations $(\ref{eq:gtt})$ and $(\ref{eq:grr})$, take the form
\begin{eqnarray}
\delta G^t_t & = & 2 \delta \nu e^{-2 \nu} \frac{\dot{R}}{R} ( \frac{\dot{R}}{R}
+ 2 \dot{\psi}) + 2 \delta \nu e^{-2 \psi} ( \frac{R'^2}{R^2} - 2 \psi'
\frac{R'}{R} + 2 \frac{R''}{R} ) \nonumber  \\
& & + 2 \delta \nu' ( \frac{\dot{R}}{R R'} e^{-2 \nu} + \frac{R'}{R}
 e^{-2 \psi}) \label{eq:gtt1} \\
\delta G^r_r & = & 2 \delta \nu\{ e^{-2 \nu} ( \frac{\dot{R}^2}{R^2} + 2 
\frac{\ddot{R}}{R}  + 2 \dot{\nu} \frac{\dot{R}}{R} ) + e^{-2 \psi} 
( \frac{R'^2}{R^2} + 2 \nu' \frac{R'}{R}) \} \nonumber  \\
& & - 2 \delta \nu' ( \frac{\dot{R}^2}{R R'} e^{-2 \nu} + e^{-2 \psi}
 \frac{R'}{R} ) \label{eq:grr1}
\end{eqnarray}
The above two relations  can be put in the E-L equation,
\begin{eqnarray}
\delta G^t_t & = & \delta T^t_t + \xi^t_t \\
\delta G^r_r & = & \delta T^r_r + \xi^r_r 
\end{eqnarray}
Using an equation of state $ p = w \rho $, and perturbing it to give $
\delta p = w \delta \rho $, the above two relations  can be put
 together , to  get expressions for the metric potential $ \delta \nu $ (and
$\delta \psi $).  
Thus
\be
\delta G^r_r = - w \delta G^t_t + w \xi^t_t + \xi^r_r
\ee
Further using  (\ref{eq:gtt1}) and (\ref{eq:grr1}) we obtain
\begin{eqnarray}
\nu' f_1(t,r) + f_2 (t,r) \delta \nu = - w \xi^t_t - \xi^r_r
\end{eqnarray}
where
\[ f_1(t,r) = 2(1+w) ( \frac{\dot{R}^2}{R R'} e^{-2 \nu} + \frac{R'}{R}
 e^{-2 \psi} )\]
and 
\begin{eqnarray}
 f_2(t,r) & = &- 2 [ e^{-2 \nu} (\frac{\dot{R}^2}{R^2} (1 + w) + 2
 \frac{\ddot{R}}{R} + 2 \dot{\nu} \frac{\dot{R}}{R} + 2 w \dot{\psi} 
\frac{\dot{R}}{R}) \nonumber \\
& & e^{-2 \psi} ( \frac{R'^2}{R^2} ( 1+ w) + 2 \frac{R'}{R}(\nu' -  w \psi')
+ 2 w \frac{R''}{R})] \nonumber 
\end{eqnarray}
Solving for $ \delta \nu $,
\be
\delta \nu(t,r) = e^{ \int \frac{f_2(t,r)}{f_1(t,r)} dr} \int e^{-\int
 \frac{f_2(t,r')}{f_1(t,r')} dr'} ( - w \xi^t_t - \xi^r_r ) dr''
\ee
Its is clear then, that  $<\delta \nu> = 0 $ as expected for the Langevin
 formalism. The two point correlations are given by
\begin{eqnarray}
<\delta \nu(t,r) \delta \nu(t',r')> & = & e^{ \int \frac{f_2(t,r)}{f_1(t,r)}
  dr + \int \frac{f_2(t',r')}{ f_1(t',r')} dr' } \int \int e^{-[\int
\frac{f_2(t,r_1)}{f_1(t,r_1)} dr_1 + \int \frac{f_2(t',r_2)}{f_1(t',r_2)}
 dr_2 ]}  \nonumber \\
& & \{ w^2 < \xi^t_t (t,r_1') \xi^t_t ( t',r_2') > + <\xi^r_r(t,r_1')
 \xi^r_r(t',r_2')> +  \nonumber \\
& & w ( <\xi^t_t(t,r_1') \xi^r_r( t',r_2')> + <\xi^r_r(t,r_1')
 \xi^t_t ( t',r_2')> ) \} \nonumber \\
& & dr_1' dr_2'
\end{eqnarray}
Putting in the noise from sec. \ref{sec:noise},
\begin{eqnarray} \label{eq:sol1}
<\delta \nu(t,r) \delta \nu(t',r')> & = & e^{ \int \frac{f_2(t,r)}{f_1(t,r)}
  dr + \int \frac{f_2(t',r')}{ f_1(t',r')} dr' } \int \int e^{-[\int
\frac{f_2(t,r_1)}{f_1(t,r_1)} dr_1 + \int \frac{f_2(t',r_2)}{f_1(t',r_2)}
 dr_2 ]}  \nonumber \\
& & \{ w^2  Cov[\rho(t,r_1'),\rho ( t',r_2') ] +
 Cov[p(t,r_1'),p(t',r_2')] \nonumber \\
& &- w ( Cov[\rho(t,r_1'), p( t',r_2')] + Cov[p(t,r_1')
, \rho ( t',r_2')] ) \}  \nonumber \\
& & dr_1' dr_2'
\end{eqnarray}
Another  solution arises from taking the other possible set of Einstein's 
tensors $\delta G^t_t  $ and $\delta G^r_r $ with the ' derivatives, replaced 
by '.' derivatives using the relation  (\ref{eq:soltr}).
\begin{eqnarray}
\delta G^t_t & = & 2 \delta \nu [ e^{-2 \nu} \frac{\dot{R}}{R} (
 \frac{\dot{R}}{R} + 2 \dot{\psi}) + e^{-2 \psi} ( \frac{R'^2}{R^2} \nonumber \\
& & - 2 \psi' \frac{R'}{R} + 2 \frac{R''}{R} ) ] + 2 \delta \dot{\nu}[
e^{-2 \nu} \frac{\dot{R}}{R} + \frac{R'^2}{R \dot{R}} e^{-2 \psi} ] \\
\delta G^r_r & = & 2 \nu [e^{-2 \nu} ( \frac{\dot{R}^2}{R^2} + 2 
\frac{\ddot{R}}{R} + 2 \dot{\nu} \frac{\dot{R}}{R} ) + \nu e^{-2 \psi}
( \frac{R'^2}{R^2} + 2 \frac{R'}{R} \nu' )] \nonumber \\
& & - 2 \delta \dot{\nu} [ \frac{\dot{R}}{R} e^{-2 \nu} + \frac{R'^2}{ R
 \dot{R}} e^{-2 \psi}]
\end{eqnarray} 
From the Einstein Langevin equations as done for the previous case, we
 obtain 
\begin{eqnarray}
 & & 2 \delta \dot{\nu}(1+w) [ \frac{\dot{R}}{R} e^{-2 \nu}  + e^{-2 \psi} 
\frac{R'^2}{R \dot{R}}  ] - 2 \delta \nu [ e^{-2 \nu} \{ 
\frac{\dot{R}^2}{R^2} (1 + w) \nonumber \\
& & + 2 \frac{\dot{R}}{R} ( \dot{\psi} w + \dot{\nu}) \} + e^{-2 \psi} (
\frac{R'^2}{R^2} ( 1 + w + 2 \frac{R'}{R} ( \nu' - w \psi') +
 2 \frac{R''}{R})] \nonumber \\
& & = -w \xi^t_t - \xi^r_r 
\end{eqnarray}
This can be written as
\be
 \delta \dot{\nu} f_3(t,r) + \delta \nu f_2(t,r) = - w \xi^t_t - \xi^r_r
\ee
where,
\be
f_3(t,r) =  2 (1+w)[ \frac{\dot{R}}{R} e^{-2 \nu} + e^{-2 \psi}
\frac{R'^2}{R \dot{R}}] 
\ee
we get a similar solution in this case, as the previous one,but in terms of 
temporal integrals. 
\be
\delta \nu(t,r) = e^{ \int \frac{f_2(t,r)}{f_3(t,r)} dt } 
\int e^{-\int \frac{f_2(t',r)}{f_3(t',r)} dt'} (- w \xi^t_t - \xi^r_r ) dt'' 
\ee
Again, as expected $<\nu(t,r)> = 0$, while the two point correlation is given
by
\begin{eqnarray}
<\delta \nu(t,r) \delta \nu(t',r')> & = & e^{\int \frac{f_2(t,r)}{f_3(t,r)} dt
+ \int \frac{f_2(t',r')}{f_3(t',r')} dt'}  \int \int e^{- \int 
\frac{f_2(t_1,r)}{ f_3(t_1,r')} dt_1 - \int 
\frac{f_2(t_2,r')}{f_3(t_2,r')} dt_2 } \nonumber \\
& & [ w^2 < \xi^t_t(t_1',r) \xi^t_t ( t_2',r')> + w( <\xi^t_t(t_1',r)
\xi^r_r(t_2',r')> + \nonumber \\
& &  <\xi^r_r(t_1',r) \xi^t_t(t_2',r')>) + < \xi^r_r(t_1',r)
 \xi^r_r(t_2',r')> ] \nonumber \\
& &  dt_1' dt_2' 
\end{eqnarray}
Putting in the noise from sec. \ref{sec:noise},
\begin{eqnarray} \label{eq:sol2}
<\delta \nu(t,r) \delta \nu(t',r')> & = & e^{\int \frac{f_2(t,r)}{f_3(t,r)} dt
+ \int \frac{f_2(t',r')}{f_3(t',r')} dt'}  \int \int e^{- \int 
\frac{f_2(t_1,r)}{ f_3(t_1,r)} dt_1 - \int 
\frac{f_2(t_2,r')}{f_3(t_2,r')} dt_2 } \nonumber \\
& & [w^2 Cov[\rho(t_1',r) \rho(t_2',r')] -w ( Cov[\rho(t_1',r),p(t_2',r')] +
 \nonumber \\
& &  Cov[p(t_1',r),\rho(t_2',r')]) + Cov[p(t_1',r),p(t_2',r')] ]
 \nonumber \\
 & &  dt_1' dt_2' 
\end{eqnarray}
Equation (\ref{eq:sol1}) and (\ref{eq:sol2}) represent two equivalent 
expressions for the two point correlation of perturbed metric potential
$\delta \nu(t,r)$, both of which show that these depend on the covariances of
 pressure and density inside the star. 
 
One could also obtain easily from equation (\ref{eq:sol1}) the case $t = t'$,
  for coincident comoving time  the two point correlator with 
dependencies on covariances of density and pressure between two different radial
coordinates, while from ( \ref{eq:sol2})
the case $ r= r'$, for  the coincident radial coordinate, with dependencies on
pressure and density at  two different comoving times can be obtained.

 We further describe the results  (\ref{eq:sol1}) and (\ref{eq:sol2})
 giving a clearer picture that emerges 
from the given expressions. An obvious question to raise here is over the 
 observable effects of the matter fluctuations or stochasticity in general. As
 mentioned earlier in the article these stochastic effects could be due to 
external 
sources, or internal dynamics of the gravitating system, but more importantly
as seen from the expressions, they show up in the form of covariances of 
pressure and density at two different points inside the collapsing cloud. 
 The pressures or density inside a collapsing star which are treated as 
 random  variables can vary differently by significant amounts,
  with respect to one another at a few points in space and time during the
 collapse.  
Specially in the context of collapsing models, this
could be significant near the critical phases of collapse, where the two
radial points are placed such that one lies inside or at the shell which is 
imploding while the other which is about to explode or outgoing, and lies at
 the outer region of the stellar configuration (similar for the temporal
 coordinates as well). One can think of stages of
 collapse just before a Supervona burst, when one of the points lies where
the core collapse would occur and the other in the outer region which would
undergo explosion. It is for a 
short period in the collasping stages or non-equilibrium 
 stages of the collapse that this could have observable effects, such that
covariances of the pressure an density at two such different points  become
 large enough.  In other words, the varying of pressure or density in one of the
regions and that in the other region (or time) near or at a critical
phase of collapse may show up very different behavior in the statistical sense
which would be captured by the terms like $ Cov[p(r,t) p(r',t)]$ and have 
significant impact on the metric perturbations.  It is important to realize
that here we are not trying to observe fluctuations of the matter content at 
a given particular point inside a star which is negligible and would not have
observable consequences, but the covariance at two different points in two very
different regions, where the matter can have significantly different 
properties, which  show up  in how the pressure and density vary with respect 
to each other at two different points. This  effect could be large enough to
 perturb the metric (while also showing up correlations in the metric 
perturbation at two points)  during the critical phases. As an analogy, one 
can compare this with complete gravitational collapse of black holes, where
interesting behaviour of quantum fluctuation is seen to emerge at  Cauchy 
horizon, that of a divergence of a regularized stress tensor as well as its
 two point fluctuations \cite{sukratu}. Of course here we are dealing with a
 classical  system,  which 
is a different picture, however in the collapsing scenario, these effects
will be more significant in observational sense than a static configuration as
 is proposed in \cite{seema}. This is because  the 
dynamical collapse  does give different regions of the massive star, 
significantly different behavior, and thus a non-equilibrium statistical
 treatment to the analysis can  be more relevant here . 
  
Thus the implications of this results lies in the fact that the two point
correlations of the induced metic fluctuations 
 act as a probe to the interior statistical properties the perfect
fluid or the matter content during critical phases of dynamical collapse.
Moreover, one can
see a drastic change in these perturbations from being negligible during smooth
 collapse phases to becoming important near or at the critical points 
of collapse. 

As a stellar configuartion can undergo many critical phases 
 during the collapse,  the above analysis can probe into the statistical
properties at each stage of criticality and compare and contrast the behavior
from the begining till the final stages of collapse, when the core settles 
down in an equilibrium configuration and these covariances become
insignificant again.  
By doing such an analysis of a collapsing stellar configuration, one could 
study detailed interior dynamics statistically. This may provide
 information about the matter content and its properties  and evolution
of the interiors through  different stages of the collapse which can be of 
Astrophysical interest. Hence such a theoretical development as discussed in
this article is more important than for a static configuration of a 
relativistic star.

 Further significance of the analysis done in this article is discussed below.
 
\section{Discussion and further directions}
  The statistical correlations of the metric perturbations shown in the 
analysis here  are the basic building blocks in an effort to  characterize the
 dynamical statistical properties, which are induced by those of the matter
 content of the interior of the star.
Thus the correlations in the metric perturbations show up the induced 
stochastic behavior.  The model treated here for the collapsing case is
 a toy model, with very basic structure, we intend to carry forward this
 formalism to more realistic situations of collapse in further work, which
 would treat different stress tensors and matter components of the interior
 for the star as well as different collapsing geometries like the  rotating 
configurations for collaspe and deviation from spherical symmetry. However the 
 method of solution of such systems, is not as simple as is treated in the
 present cse, but will involve developments on the lines of theory of
 perturbed system for non-radial collapse using different gauges. 
The most important challange is to work out formalism on the same lines
 without using the
 Cowling approximation, as we need to see the effects of the matter field 
stochasticity on the spacetime. Though it may seem 
initially, that these effects would be minor and not of much Astrophysical
 interest, we expect to see relevant results,  in cases, where one is 
 interested in critical stages of collapse in compact relativistic objects
 or configurations, for both radial as well as non-radial perturbations.
The non-equilibrium statistical physics developed for such applications would
 be of interest to dynamical collapse of gravitating bodies and capture and 
characterize the interiors in the language of spacetime perturbations. 
It is likely  that near the end states of collapse, this may enable one to
study details of the behavior of geometry and affect results and parameters at
 horizons of interest, viz the event horizon in case of formation of black
 holes or affecting the bounce models of collapse and relevant parameters 
therein.  It would be of interest to examine if boundary conditions  for the 
collapsing cloud are affected or modified by including the stress tensor
 fluctuations for different cases of imperfect, dissipative fluid with heat
 flux and so on, specially when the matter fluctuations in the crust 
of a relativistic star are significant. Non equilibrium effects may become
very interesting for such cases, which is an important research direction that
can follow from the work presented in this article.  
\section*{Acknowledgements}
Seema Satin is thankful to Sukanta Bose for useful discussions and Bei Lok Hu
 for relevant directions.  . The work
 carried out in this paper has been funded by Department of Science and 
Technology (DST) India, WoS-A fellowship, grant no. 
DST/WoS-A/2016/PM100. 

\end{document}